\documentclass[reprint,twocolumn]{revtex4}
\usepackage{graphicx}
\usepackage{amsmath}

\begin{document}

\title{Divergence of an orbital-angular-momentum-carrying beam upon propagation}
\author{Miles Padgett$^1$}
\author{Filippo M. Miatto$^2$}
\author{Martin Lavery$^1$}
\author{Anton Zeilinger$^{3,4}$}
\author{Robert W. Boyd$^{1,2,5}$}
\affiliation{$^1$School of Physics and Astronomy, University of Glasgow, Glasgow G12 8QQ, United Kingdom}
\affiliation{$^2$Dept. of Physics, University of Ottawa, 150 Louis Pasteur, Ottawa, Ontario, K1N 6N5 Canada}
\affiliation{$^3$Institute for Quantum Optics and Quantum Information (IQOQI), Austrian Academy of Sciences, Boltzmanngasse 3, A-1090 Vienna, Austria}
\affiliation{$^4$Vienna Center for Quantum Science and Technology, Faculty of Physics, University of Vienna, Boltzmanngasse 5, A-1090 Vienna, Austria}
\affiliation{$^5$The Institute of Optics, University of Rochester, Rochester, New York 14627, USA}

\date{\today}
\begin{abstract}
There is recent interest in the use of light beams carrying orbital angular momentum (OAM) for creating multiple channels within free-space optical communication systems.  One limiting issue is that, for a given beam size at the transmitter, the beam divergence angle increases with increasing OAM, thus requiring a larger aperture at the receiving optical system if the efficiency of detection is to be maintained.  Confusion exists as to whether this divergence scales  linarly with, or with the square root of, the beam's OAM. We clarify how both these scaling laws are valid, depending upon whether it is the radius of the Gaussian beam waist or the rms intensity which is kept constant while varying the OAM.
\end{abstract}

\maketitle

\section{introduction}
Over the past 20 years there has been a growing interest in the orbital angular momentum (OAM) of light, which is carried by any optical beam possessing helical phase fronts \cite{AlisonMiles}.  For phase fronts described by an azimuthally dependent phase factor $\exp(i\ell\phi)$, the OAM is equivalent to $\ell\hbar$ per photon \cite{PhysRevA.45.8185}.  This OAM of a light beam can be derived in a number of ways, but for our purposes it can be seen to arise from the local direction of the Poynting vector.  At all points within the beam the Poynting vector is perpendicular to the phase fronts and hence possesses an azimuthal momentum component \cite{Padgett199536}.  As confirmed by measurement \cite{Leach:06}, helical phase fronts lead to the local Poynting vector being skewed with respect to the beam axis by an angle $\beta$, which is a function of $r$, the radial distance from the beam axis and $k_0=2\pi/\lambda$:
\begin{align}
\beta=\frac{|\ell|}{k_0r},
\end{align}

One analytically simple, and widely used, form of  helically-phased beams is the Laguerre-Gaussian (LG) modes having a field amplitude given as the product of a Gaussian term with beam waist $w_0$, (defined as the radius at which the electric field of the Gaussian term is lower by a factor $1/e$), a generalised Laguerre polynomial with indices $|\ell|$ and $p$, and  the coordinate $r$ raised to the power $|\ell |$ (see Eq. (3)).

In this work we restrict ourselves to the paraxial regime and small numerical aperture where the size of any spatial features in the modes or their superpositions is significantly larger than the optical wavelength.  For $p=0$, the LG modes are single-ringed annular modes with zero on-axis intensity and an azimuthal phase term $\exp(i\ell\phi)$, unless also $\ell$ is zero, in which case one has the fundamental Gaussian mode. The radius of maximum optical intensity of one of these modes, $r(I_\mathrm{max})$, is given as \cite{Padgett199536}
\begin{align}
r(I_\mathrm{max})=\sqrt{\frac{|\ell|}{2}} w(z),
\label{imax1}
\end{align}
where $w(z)=w_0\sqrt{1+z^2 / z_R^2}$ and $z_R=\frac{ 1}{2}   kw_0^2$ is the Rayleigh range (here $k=2\pi/\lambda$).  One notes that for high OAM values the size of the LG mode is much larger than the beam waist. The restriction of our study to the paraxial regime also implies that  $w(z) \gg \lambda$ and hence that $\beta$  (see Eq.\ (1)) remains small for all radii at which the beam has a non-negligible intensity.

In terms of their complex amplitudes, LG beams with different mode indices are orthogonal to each other and this suggests their use in communication applications where the orthogonality of the modes creates independent channels.  A simple optical communication system based on OAM was demonstrated over 10 years ago \cite{Gibson:04}, but recently the idea has been revisited with reports of both high data rates \cite{Wilner:12} and outdoor demonstrations \cite{Krenn:14}.  One issue of critical importance is the degree to which the use of OAM-carrying beams require the use of larger aperture optics.  Confusion exists as to whether the far-field divergence of a beam carrying OAM scales linearly or with the square root of its angular momentum.

\section{divergence of a beam}
The skew angle of the Pointing vector calculated with respect to the beam axis might be thought to approximate the angular divergence of the beam.  However, an additional contribution to the divergence that has to be considered is the normal diffractive spreading arising from the finite beam diameter.  Rather than the beam waist $w_0$, or the radius of maximum intensity $r(I_\mathrm{max})$, the divergence of a light beam is governed by the standard deviation of its spatial distribution, $r_\mathrm{rms}$ \cite{Stelzer200051,miles}. 
The calculation of these quantities is  straightforward. We start from the definition of a LG mode from which we find that the intensity distribution for the lowest-order radial mode ($p=0$) is given by
\begin{align}
I_\ell(r,\phi,z)= \frac{2}{w(z)^2\pi|\ell|!}\left(\frac{\sqrt{2}r}{w(z)}\right)^{2|\ell|}\exp\left(-\frac{2r^2}{w(z)^2}\right),
\end{align}
which is normalized such that $\int I_\ell(r,\phi,z)\, rdrd\phi=1$.
From this, we can calculate the radial position of the maximum of intensity, by solving the equation ${\partial I_\ell}/{\partial r}=0$, obtaining (as anticipated in eq.~\eqref{imax1} \cite{Padgett199536})
\begin{align}
\label{imax}
r(I_\mathrm{max})=\sqrt{\frac{|\ell|}{2}}w(z),
\end{align}
However, as $r(I_\mathrm{max})$ is zero for $\ell=0$, clearly one cannot generally use this quantity to derive the angular spread of the beam as it propagates. As has been reported previously, a better quantity is the standard deviation of the spatial distribution of the beam, $r_\mathrm{rms}$, which is a function of $z$ and it is given by the square root of the radial variance of the intensity distribution \cite{Phillips:83}:
\begin{align}
r_\mathrm{rms}(z)&=\sqrt{2\pi{\int_0^\infty}r^2I_\ell(r,\phi,z)\,rdr}\\
&=\sqrt{\frac{|\ell|+1}{2}}w(z),
\label{rrms}
\end{align}

In the plane of the beam waist, this simplifies to
\begin{align}
r_\mathrm{rms}(0)=\sqrt{\frac{|\ell|+1}{2}}w_0,
\label{rrmswaist}
\end{align}

\begin{figure}[!h]
\begin{center}
\includegraphics[width=0.9\columnwidth]{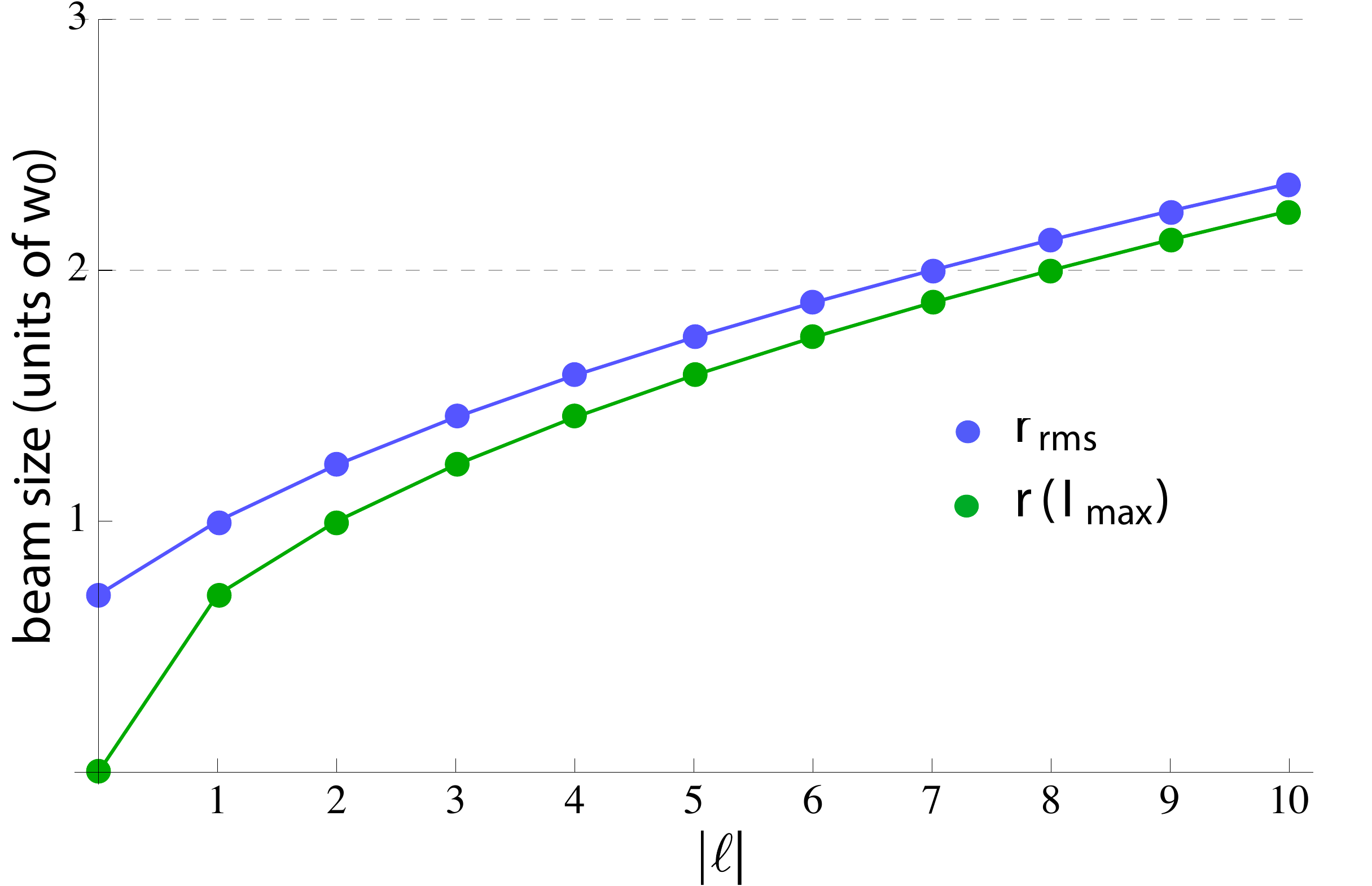}
\caption{\label{fig1}A comparison between the radius in the pupil plane of maximum intensity and the rms radius of the spatial distribution of a $p=0$ LG mode. For large $\ell$, the two radii get asymptotically close, as $r(I_\mathrm{max})$ is proportional to $\sqrt{|\ell|}$, while $r_\mathrm{rms}$ is proportional to $\sqrt{|\ell|+1}$. }
\end{center}
\end{figure}

Fig.~\ref{fig1} shows the $\ell$ dependence of $r(I_\mathrm{max})$ and $r_\mathrm{rms}(0)$ in the pupil plane.  One notes that
\begin{align}
\frac{r_\mathrm{rms}}{r(I_\mathrm{max})}=\sqrt{\frac{|\ell|+1}{|\ell|}},
\end{align}
which tends to 1 as $|\ell|$ increases, so these two radii become equal for large $\ell$, see Fig.~\ref{fig1}.

Given the radius $r_\mathrm{rms}(z)$ as a function of propagation distance, $z$, we can calculate the corresponding divergence angle of a $p=0$ LG mode, $\alpha_\ell$, to be
\begin{align}
\alpha_\ell(z)&=\arctan\left(\frac{\partial r_\mathrm{rms}(z)}{\partial z}\right)\\
&= \arctan\left(\sqrt{\frac{|\ell|+1}{2}}\frac{w_0^2}{z_R^2}\frac{z}{w(z)}\right).
\end{align}
This divergence angle is a function of $z$, but it asymptotically reaches a limiting value (when the beam is sufficiently far from the waist). Within the paraxial regime, for small $\alpha_\ell$, we can equate the tangent of the angle with the angle and for $\lim_{z\rightarrow\infty} z/ w(z)=z_R / w_0$, so in the far field, the divergence angle of a $p=0$ LG mode is given by
\begin{align}
\label{alpha}
\alpha_\ell=\sqrt{\frac{|\ell|+1}{2}}\frac{\lambda}{2 \pi w_0}.
\end{align}

\begin{figure}[!h]
\begin{center}
\includegraphics[width=0.9\columnwidth]{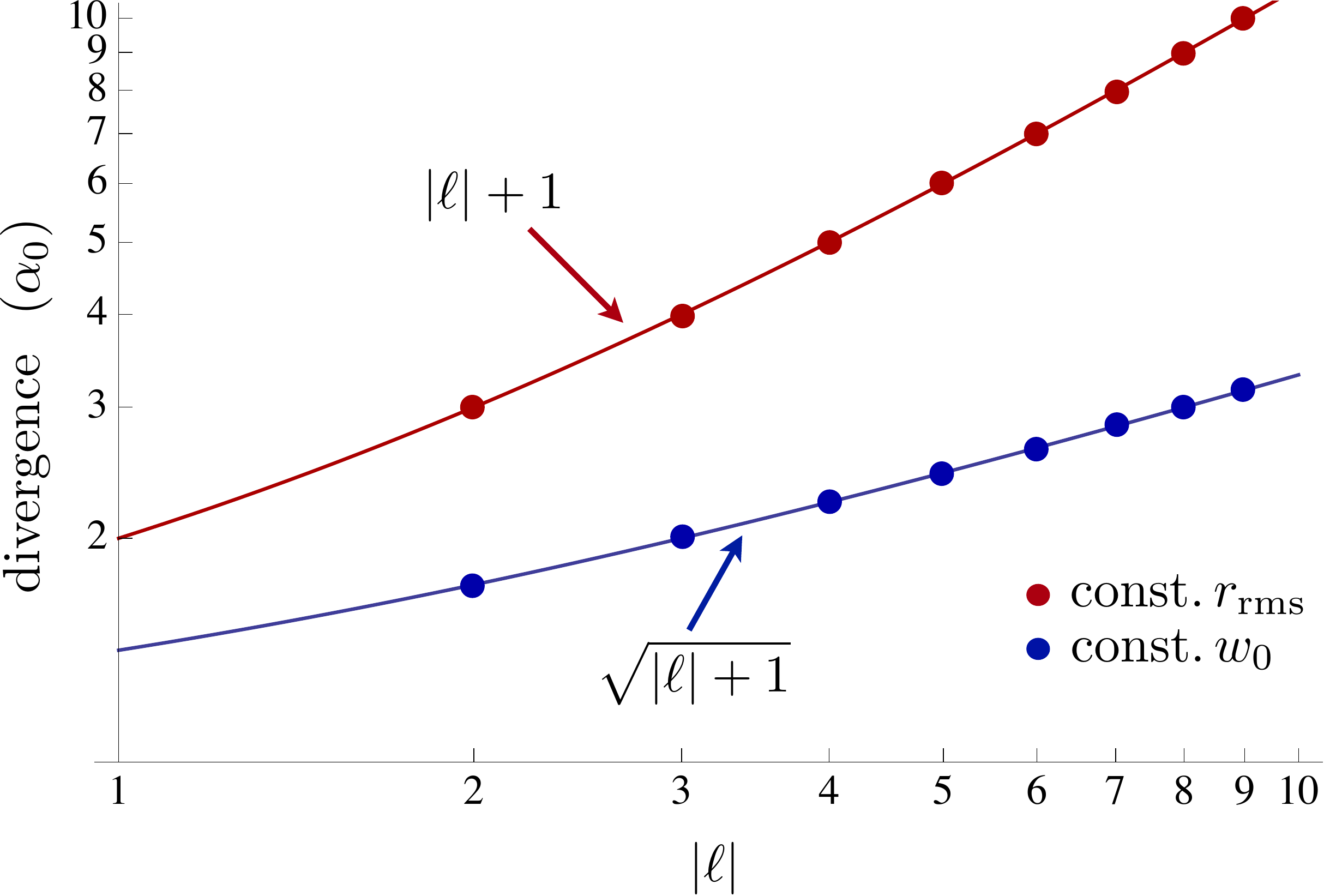}
\caption{The divergence of an OAM beam obtained by keeping the $r_\mathrm{rms}$ radius in the pupil plane constant is higher than for an OAM beam obtained by keeping the Gaussian waist $w_0$ constant. Moreover, the first scales as $|\ell|+1$ whereas the second scales as $\sqrt{|\ell|+1}$. This log-log plot is given in units of the divergence angle $\alpha_0=w_0/\sqrt{2}z_R$ of a fundamental Gaussian beam. The lines serve purely as a guide to the eye. \label{fig2} }
\end{center}
\end{figure}

Eq.~\eqref{alpha} is a convenient form for expressing the $\ell$ dependence of the beam divergence whilst holding $w_0$ constant, revealing the approximate square root scaling, see figure \ref{fig2}. Alternatively the divergence angle can be written in terms of other quantities, for instance, using Eq.~\eqref{rrmswaist}, one can instead express the $\ell$ dependence of the beam divergence whilst holding $r_\mathrm{rms}$ constant, as   
\begin{align}
\label{linear}
\alpha_\ell=\frac{|\ell|+1}{k_0r_\mathrm{rms}(0)},
\end{align}
Under this condition $w_0$ itself becomes a function of $\ell$ and in fact scales as $ (|\ell | + 1)^{-1/2}$.  In this case, corresponding to a fixed $r_\mathrm{rms}$, we see that the beam divergence scales linearly with $\ell$, see figure \ref{fig2}.

\section{Discussion}

Typical of a practical system is that the aperture of the launch optics is fixed by design, with a limiting radius $R$.  Since the size of the LG beam can be much bigger than $w_0$, if wishing to design a system incorporating a wide range of different OAM modes with a constant beam waist, then one must choose a waist $w_0\ll R$.  LG beams of a fixed beam waist are produced by a cylindrical-lens mode converter  \cite{Beijersbergen1993123}.  Using a mode converter of this type to produce the modes one would observe the square root scaling of the far-field beam divergence \eqref{alpha}.

A more frequently used and flexible method of producing beams carrying OAM is the illumination of a diffractive optical element with an expanded Gaussian beam from a conventional laser.  If the diffractive element is a spatial light modulator (or similar) then the displayed kinoform can be updated to change the beam type allowing a switching of the system from channel to channel.  Most common of such kinoforms is a forked diffraction grating that produces a helically-phased beam \cite{Soskin:03}.  In the plane of the element, the resulting beam has a Gaussian intensity distribution but with a helical phase structure, corresponding to a superposition of different LG modes all of the same $\ell$ but various values of  $p$. The precise $p$-weighting of this superposition is a complicated function of both $|\ell|$ and the value of beam waist $w_0$ chosen to perform the decomposition.  Rather than describing the resulting beam as having a particular beam waist, the invariant quantity is simply the size of the illumination beam and hence  $r_\mathrm{rms}(0)$, which cannot exceed the radius of the launch optics. Such a system then exhibits a far-field beam divergence better described by eq.~\eqref{linear}, i.e. a scaling approximately proportional to $|\ell|$. This linear scaling of the beam size has previously been noted within optical tweezers and microscope systems \cite{Curtis:03}.

We note that the reasoning we have followed is compatible with an understanding of the scaling in terms of the spatial resolution of an optical system.  For example, if one considers a transmitted beam comprising a superposition of $\pm\ell$ modes, then the resulting beam profile is an annular ring of $2|\ell|$ maxima, or ``petals''.  In the far field the spacing between these petals cannot be smaller than the resolving power of the optical system.  Since the number of petals scales with $|\ell|$ this implies that the beam size and hence the beam divergence must scale also with $|\ell|$.  A square root scaling of the beam divergence $\alpha_\ell$ and corresponding decrease in petal separation would seemingly be in conflict with this simple resolution argument.  However, a square root scaling requires a constant beam waist $w_0$ which implies a beam size $r_\mathrm{rms}$ in the waist plane and hence a transmission aperture which scales with the square root of $|\ell|$.  This increasing aperture itself supports a higher spatial resolution, meaning that the reduced spacing of the resulting petals remains compatible with the resolution limits.

\section{conclusions}
In conclusion, we have explained how optical systems for the propagation of light beams carrying OAM can exhibit a far-field beam divergence that scales either with the square root of the OAM or linearly with the OAM. The square root scaling applies to systems where the beam waist of the Gaussian term, $w_0$, is held constant for all values of $\ell$, such as the case where the beams are produced using a cylindrical lens mode converter \cite{Beijersbergen1993123}. The linear scaling applies to systems where the rms radius of the launch beam is held constant, such as the case where the beam is produced by a simple forked diffraction grating implemented on a spatial light modulator illuminated with a gaussian source of fixed beam waist \footnote{One notes that more sophisticated hologram designs can produce arbitrary combinations of LG modes with any values of $\ell$, $p$ and $w_0$ (see for instance M. Dennis et al. \emph{Nat. Phys.} \textbf{6}, 118 (2010)), albeit at greatly reduced efficiency compared to the simple fork design that is widely used}.

\bibliography{divergencedraft}{}
\bibliographystyle{unsrt}

\end{document}